\documentstyle[prl,aps,epsf,epsfig,multicol]{revtex}

\epsfclipon

\newcommand{\be}{\begin{equation}}
\newcommand{\ee}{\end{equation}}
\newcommand{\bes}{\begin{eqnarray}}
\newcommand{\ees}{\end{eqnarray}}
\newcommand{\bma}{\left( \begin {array}}
\newcommand{\ema}{\end {array} \right)}

\newcommand{\C}{\langle C\rangle}

\begin{document}

\title{Opacity and Entanglement of Polymer Chains}

\author{Peter Grassberger}

\address{John-von-Neumann Institute for Computing, Forschungszentrum J\"ulich, D-52425 J\"ulich, Germany}

\date{\today}

\maketitle
\begin{abstract}  
We argue that the mean crossing number of a random polymer configuration 
is simply a measure of opacity, without being closely related to entanglement as 
claimed by several authors.
We present an easy way of estimating its asymptotic behaviour numerically. These 
estimates agree for random walks (theta polymers), self-avoiding walks, and for compact globules 
with analytic estimates giving $\log N, a-b/N^{2\nu-1},$ and $N^{1/3}$, respectively,
for the 
average number of crossings per monomer in the limit $N\to \infty$. While the 
result for compact globules agrees with a rigorous previous estimate, the result 
for SAWs disagrees with previous numerical estimates.

  \vspace{4pt}
\end{abstract}

\begin{multicols}{2}

Topological properties of linear polymer chains are, strictly spoken, only defined for 
closed rings where the theory of knots can be applied with several non-trivial 
results \cite{orlandini-etal,kholodenko-vilgis,moroz-kamien}. But even for open 
chains quantities like linking, twist and writhe can be applied. Important applications
are supercoiling of DNA, DNA electrophoresis \cite{stasiak},
and the supposed absence of knots in protein backbones
\cite{kholodenko-vilgis}. Another class of problems where entanglement should play 
a crucial (but so far not yet fully understood) role is the rheology of semidilute 
solutions.

One important quantity for studying entanglement is the {\it writhe} which is defined 
as the number if signed crossings in a projection of a 3-d non-selfintersecting 
oriented curve. 
If two parts of the curve seem to cross when seen from a particular angle, this 
crossing contributes $+1$ or $-1$ to the writhe, depending on whether the direction of 
the front part is obtained by a right or left turn from the direction of the part 
behind it. Its interest stems from the fact that for closed loops it is 
related to linking which is a topological invariant \cite{moroz-kamien}. 

The {\it number of crossings} $C$ was introduced in \cite{rensburg,orlandini-etal} as 
a simplified version of the writhe. In it, all crossings contribute with the same sign.
Of particular interest is its average value $\C$, averaged over all 
angles of projection, called the {\it mean crossing number}.  In these papers it 
was shown that for a self avoiding random walk 
(SAW) of $N$ straight bonds
\be
      \C \sim N^\alpha    \quad {\rm for} \;\; N\to\infty
\ee
with $1\le \alpha \le 2$. Numerical simulations gave $\alpha=1.122\pm 0.005$, but it 
was argued that this might actually be a lower estimate, the true value being higher
\cite{orlandini-etal}.

In later simulations, Arteca \cite{arteca1} found a value $1.20\pm 0.04$ for SAWs 
and $1.34$ to $1.4$ for protein backbones \cite{arteca2}. Indeed an increase of the 
value of $\C$ had also been seen in \cite{orlandini-etal} 
for SAWs with self-attraction, and it was conjectured in 
\cite{arteca2,arteca3} that $\C$ is a useful observable for 
detecting the coil-globule transition. 
Due to its supposed 
importance, $\C$ was called the ``entanglement complexity" in 
\cite{kholodenko-rolfsen}, and was shown there (by non-rigorous arguments) to be 
$< 1.4$ for random configurations.

It is the purpose of this note to show that $\C$ can be easily
estimated by using well
known formulae for generic intersections of random fractals \cite{falconer}.
Take two fractal sets $X$ and $Y$ with dimensions $D_X$ and $D_Y$, embedded in a
space of dimension $d$. Then, their intersection has dimension $D_{X\cap Y} =
D_X + D_Y - d$ for nearly every relative position and orientation, provided 
this intersection is non-empty. In the
present case, $X$ is the curve to be studied, $Y$ is a line of view (therefore
$D_Y=1$) which passes through $X$, and $d=3$. This gives
\be
   D_{X\cap Y} = D_X -2.
\ee
If this is positive, the average number of intersections between the line of 
view and $X$ increases as $m\sim N^{D_{X\cap Y}/D_X}$ for $N\to \infty$. 
Actually, for this to be true we either have to assume that $X$ is a true 
fractal without lower length cutoff (which is not true for random walks with 
finite step size $a$), or we have to fatten $Y$. Thus we consider instead of 
a single line of view a cylinder whose thickness is of the order of the 
step size $a$ (the precise value is irrelevant), and the above number of intersections 
has to be interpreted as the number of crossings between projected bonds 
within a distance $O(a)$. This scales in the same way as the number of 
crossings per bond. Thus we obtain immediately
\be
   \C/N \sim N^{\tilde{\alpha}}
\ee
with
\be
  \tilde{\alpha} =\alpha-1 = D_{X\cap Y}/D_X = 1-2/D_X.           \label{alpha}
\ee
This is the case for compact (collapsed) polymers where $D_X=3$ and therefore
\be
  \tilde{\alpha} = 1/3 \qquad {\rm (compact \ globules)},
\ee
in perfect agreement with the exact results of \cite{cantarella,buck} and with
the numerical values for protein backbones.

This argument has to be modified when $D_X \le 2$. For $D_X<2$ (i.e., for SAWs
where $D_{SAW}=1/\nu$ with $\nu=0.5877$ \cite{li-madras-sokal}), the above argument
gives, mutatis mutandis, not the leading term but the first subleading correction.
The leading term is $\C/N\sim const$ since $\C/N$ cannot vanish for $N\to\infty$
\cite{rensburg,orlandini-etal}. The average number
of intersections per bond scales now as
$m \sim const - N^{D_{X\cap Y}/D_X}$. This gives
\be
   \C/N \sim a_0 - a_1/N^{\tilde{\alpha}}   \qquad {\rm (SAW)}          \label{SAW}
\ee
with $\tilde{\alpha}$ given by Eq.(\ref{alpha}).

For $D_X = 2$, finally, we expect logarithmic dependence. This is the case for
ordinary random walks (RW) and for $\Theta$-polymers (actually, since we want no true
self-intersections, we should consider here only the latter. But we shall speak 
of RWs for simplicity. Later, in eq.(\ref{RW}), we will also allow true RWs). 
The number of times a projection of an $N$-step RW 
comes back to a previously visited site, within a finite distance $a$, increases 
$\sim \ln N$ \cite{feller}, and therefore 
\be
   \C/N \sim \ln N\;.    \qquad {\rm (RW)}        \label{RW-0}
\ee

In order to verify Eqs.(\ref{SAW},\ref{RW-0}) numerically, we performed Monte Carlo
simulations (we did not make simulations for the collapsed case since there the
theoretical result is too obvious). In these simulations we indeed did not 
calculate $\C$ proper, but a closely related quantity which should
show the same asymptotic behaviour and is much easier to calculate numerically.

We consider walks on a simple cubic lattice, and consider only projections along 
one of the 3 axes. We consider the sites $i$ on the projection plane which are 
visited $m_i$ times with $m_i>1$, and count the number of pairs of visits to the 
same site. Dividing this by $N+1$, we obtain
\be
    B = {1\over 2(N+1)} \sum_i \,(m_i-1)m_i
\ee
where $i$ runs over all sites the plane. We call this the 
``opacity". It can obviously also be defined for ordinary (i.e. non-self avoiding)
walks. If there are no double visits in 
the projection, the opacity is zero, while it diverges with $N$ for a compact 
object whose thickness along the line of view diverges with $N$.

The main differences of $B$ with respect to $\C$ are that we do not require 
transversality of the crossings and do not make 
any angular averaging. Indeed we project along atypical directions where 
bonds would not intersect transversally but overlap, so that $C$ cannot be 
properly defined. As a consequence, the numerical calculation of $B$ is trivial 
in comparison with the calculation of $\C$, for any given 
configuration. Nevertheless, we conjecture that both show the same asymptotic 
behaviour. Indeed, if this were not the case (i.e., if $C$ would depend strongly 
on the angle of view), $\C$ would be an average over a strongly
fluctuating quantity and would presumably not be of much practical use. But there 
exists strong evidence that projections and intersections of random fractal 
objects show generic features independent of the angle of projection resp. 
intersection. Finally, our theoretical discussion used generic projections
and applies therefore strictly spoken only to $\C$. If we find 
agreement for $B$, this suggests that the arguments are correct 
for $\C$ a fortiori.

\begin{figure}[b]
  \begin{center}
    \psfig{file=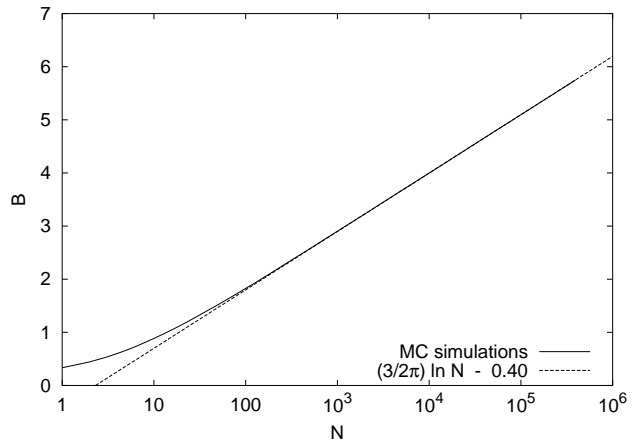,width=6.cm,angle=270}
    \vglue0.2cm
    \begin{minipage}{8.5cm}
      \caption{Log-linear plot of $B$ versus $N$ for ordinary random walks 
       (full line). The dashed line shows the analytic prediction Eq.(\ref{RW}),
       with an additive constant not predicted by theory. The value of this 
       constant obtained by fitting the data is $-0.398\pm 0.03$.
      }
  \end{minipage}
\end{center}
\label{fig1}
\end{figure}

\begin{figure}[b]
  \begin{center}
    \psfig{file=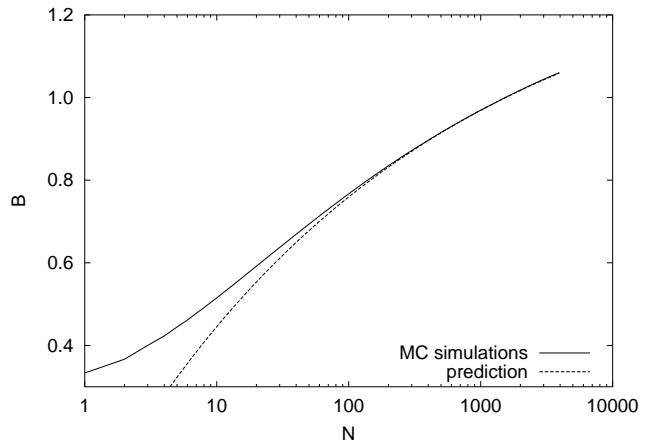,width=6.cm,angle=270}
    \vglue0.2cm
    \begin{minipage}{8.5cm}
      \caption{Log-linear plot of $B$ versus $N$ for SAWs
       (full line). The dashed line shows the analytic prediction Eq.(\ref{SAW}),
       where we have fitted numerically the constants $a_0=1.39$ and $a_1=1.415$.
      }
  \end{minipage}
\end{center}
\label{fig2}
\end{figure}

For RW, we can indeed estimate $B$ more precisely, and give the prefactor in the 
analogon to eq.(\ref{RW-0}). A projection of an $N$-step lattice RW along one 
of the coordinate axes is a RW of $2N/3$ steps
in the plane, for large $N$. The number of distinct sites visited by the latter is
$\approx (2\pi N/3)/\ln N$ \cite{feller}. Thus $m\approx (3/2\pi) \ln N $, and
\be
   B \sim {3\over 2\pi} \ln N\;.    \qquad {\rm (RW)}        \label{RW}
\ee
Results from $10^4$ RWs of $4\times 10^5$ 
steps each are shown in Fig.1, together with the prediction Eq.(\ref{RW}). 
Adding an offset which we had not tried to calculate analytically and which turned 
out to be precisely $-0.4$ within the estimated error bars, we 
find perfect agreement. 

For simulating SAWs we used the PERM algorithm \cite{grass} with Markovian 
anticipation bias \cite{causo}. This sample contained $5\times 10^6$ SAWs
(of which ca. 150,000 were strictly independent) of length $4000$. 
Results are shown in Fig.2. There we show also the
curve $1.39-1.415/N^{0.1754}$ which obviously gives a perfect fit for large $N$, 
verifying Eq.(\ref{SAW}). We should point out that our numerical values are 
in surprisingly close agreement with those shown in Fig.3 of \cite{orlandini-etal},
given the fact that we do not measure exactly the same quantity. In particular, 
if we would make a least square fit to our data with $400<N<1500$, we would also 
get $\alpha \approx 1.1$, in rough agreement with \cite{orlandini-etal}. But clearly 
such a fit would have a disastrous chi-squared. This clearly suggests that corrections 
to normal scaling have been mistaken in \cite{orlandini-etal,arteca1} for an 
anomalous power law.

Up to now we have only considered walks on the simple cubic lattice. Off-lattice 
polymers can be treated in the same way, by replacing the condition of exact coincidence
by an approximate one. Two monomers $i$ and $k$ contribute then to $B(\hat{n})$, where 
$\hat{n}$ is the direction of projection, if $|(\vec{x}_i-\vec{x}_k) \hat{n}| < 
\epsilon$ for some suitably chosen accuracy $\epsilon$. Notice that the scaling 
behaviour of $B(\hat{n})$ should not depend on $\epsilon$. We should add finally 
that $B$ can also be evaluated for any set of sites, not necessarily being lined up to 
form a toplogically linear chain. We can use it therefore to measure opacities of 
branched polymers, vesicles, clusters, or droplets.

Although we have not proven rigorously that $B$ scales in the same way as 
$\C/N$, we believe that there is little room to doubt it. In any case,
neither $\C$ nor $B$ are proper measures of entanglement, since 
e.g. the same scaling laws as for open chains should be observed also for unknotted 
loops. Instead, both are measures of opacity. Of course they can be used to 
monitor the theta transition from an open coil to a collapsed globule, since the 
transparency of a coil decreases during the collapse. But it 
is not clear what advantage they offer compared, e.g., to the gyration radius. 
The large corrections to the asymptotic behaviour seen in Fig.2 (which have previous
authors even mislead to postulate anomalous scaling laws) should be a warning that 
the interpretation of numerical values might be difficult sometimes. In any case, 
since the calculation of $B$ is much simpler than that of $\C$
while it gives basically the same information, it should be prefered in any application.

I am indebted to Walter Nadler and Rodrigo Quian Quiroga for very useful and stimulating 
discussions, and to Stuart Whittington for pointing out an error in the original manuscript.

\end{multicols}

\end{document}